\documentclass{iau_FM}
\usepackage{graphicx}

\title[IAU 2015 FM 18.~~Galaxy Group Scaling Relations] 
{Galaxy Goup Scaling Relations}

\author[R. Brent Tully]   
{R. Brent Tully$^1$}

\affiliation{$^1$Institute for Astronomy, University of Hawaii\\
2680 Woodlawn Drive, Honolulu, HI 96822, USA\\
email: {\tt tully@ifa.hawaii.edu} \\}

\pubyear{2015}
\jname{IAU Astronomy in Focus} 
\editors{Edith Falgarone \& Bruce Elmegreen, ed.}
\begin{document}

\maketitle

\begin{abstract}
What is a galaxy group?
\keywords{Galaxies: groups; mass function}
\end{abstract}


This brief account summarizes a lecture on group scaling relations that itself was a summary of two recent publications (Tully 2015$a,b$).  The overarching motivation was to build a group catalog.   Such catalogs are subject to biases created by selection criteria.  It is an obvious concern that we access fewer group member  with increasing distance.  The relationship between luminosity and mass changes with group properties.  Most fundamentally, the dimensions and velocity dispersions of groups scale with mass, so it is quite inappropriate to fix these parameters at single values in attempting to identify groups.

In Tully(2015$a$) the goal was to establish the size and velocity dispersion properties of groups over 3 decades in mass, from $10^{12}$ to $10^{15}~M_{\odot}$.  It was demonstrated that the second turnaround radius, $R_{2t}$, is an observable proxy for the virial radius.  Velocity dispersions, $\sigma_p$, are determined from spectroscopy of galaxies within this radius and masses, $M_v$, are determined from the virial theorem.

The identification of group members is often messy.  However it is possible to isolate clean cases, for example, by choosing groups that project against voids.  Candidate groups were chosen to span a wide range in galaxy numbers and types.  Wide fields encompassing target groups were observed with the Canada-France-Hawaii and Subaru telescopes.  The smallest groups to be studied, those with the most limited memberships, lie sufficiently nearby that their constituents can be resolved with Hubble Space Telescope imaging.  These observations provided accurate distances, hence unambiguous membership identifications.

The observations were designed to test theoretical expectations.  In the approximation of spherical collapse, any specified phase of collapse depends on the inverse square root of local density.  Focusing on the specific phase of second turnaround, the relation $t_{today} \sim \rho_{2t}^{-1/2}$ implies scaling relations between the two observables $R_{2t}$ and $\sigma_p$ and the virial radius $M_v$.

The main result of the first paper was the unambiguous demonstration that the anticipated scaling relations are, in fact, seen.  The products of interest are the coefficients of the fits.  There are two independent relations and one derivative relation:
$$R_{2t} = 0.215 (M_{12})^{1/3} h_{75}^{-2/3} ~{\rm Mpc}$$
$$\sigma_p/R_{2t} = 368 h_{75} ~{\rm km/s}$$
$$M_v = 2.0\times10^6 \sigma_p^3 h_{75}^{-1}~M_{\odot}$$
where $M_{12} = M_v/10^{12}~M_{\odot}$.  In this study, distances were directly measured but the scale is compatible with H$_0 = 75$ km/s/Mpc.  The observed correlations are seen in Figure 1. 

\begin{figure}[]
\begin{center}
\includegraphics[width=1.7in]{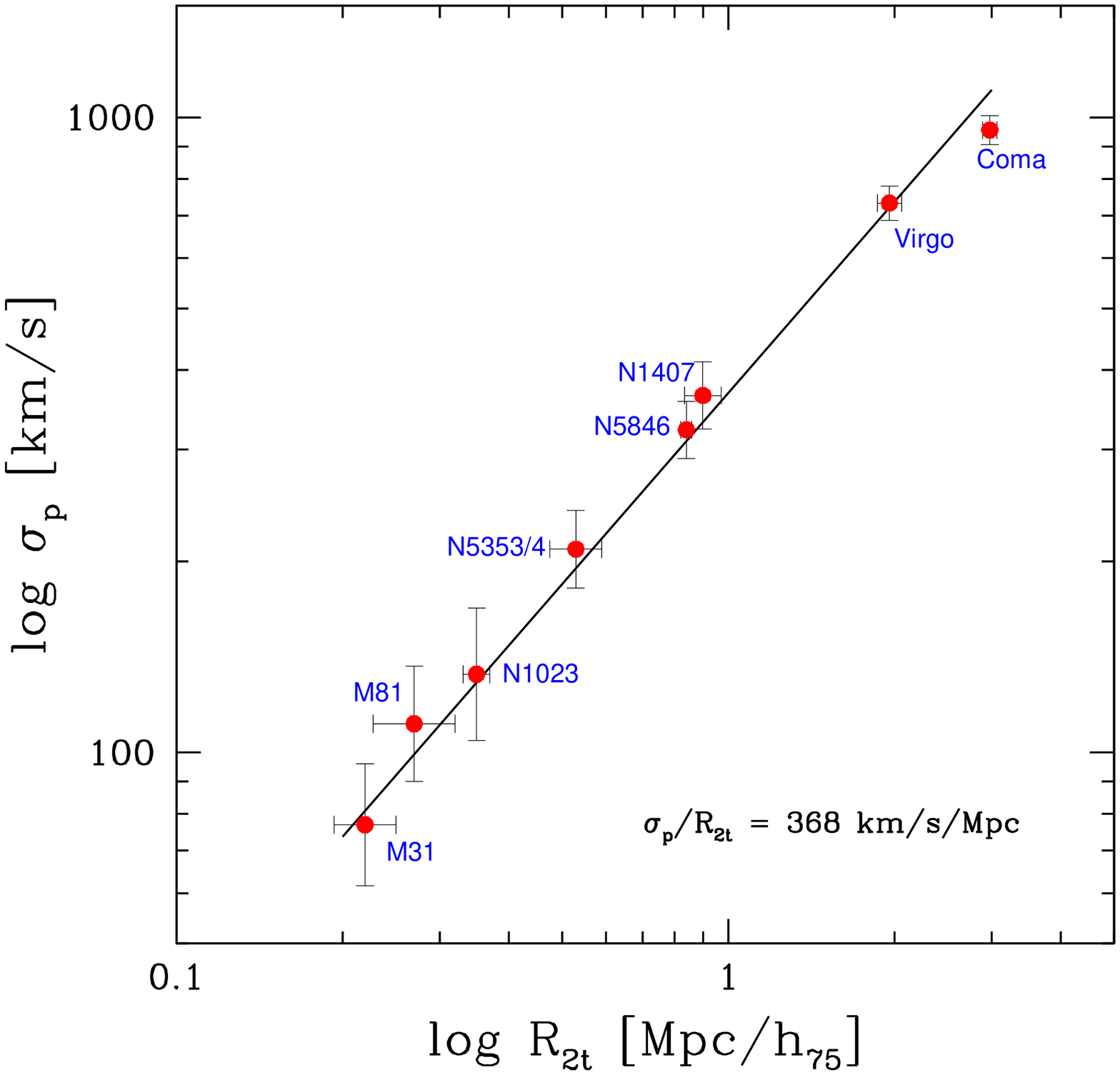} 
\includegraphics[width=1.7in]{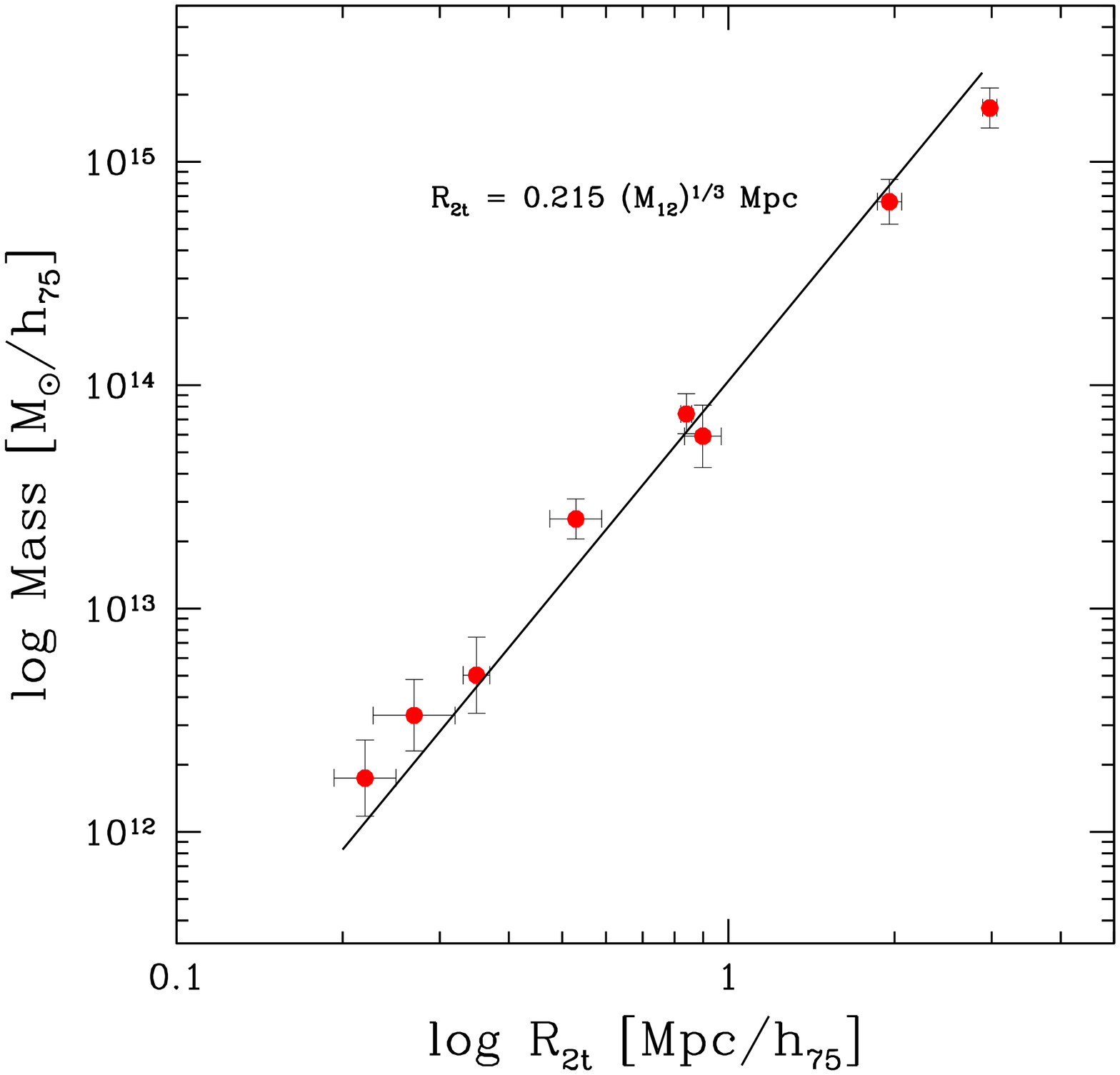} 
\includegraphics[width=1.7in]{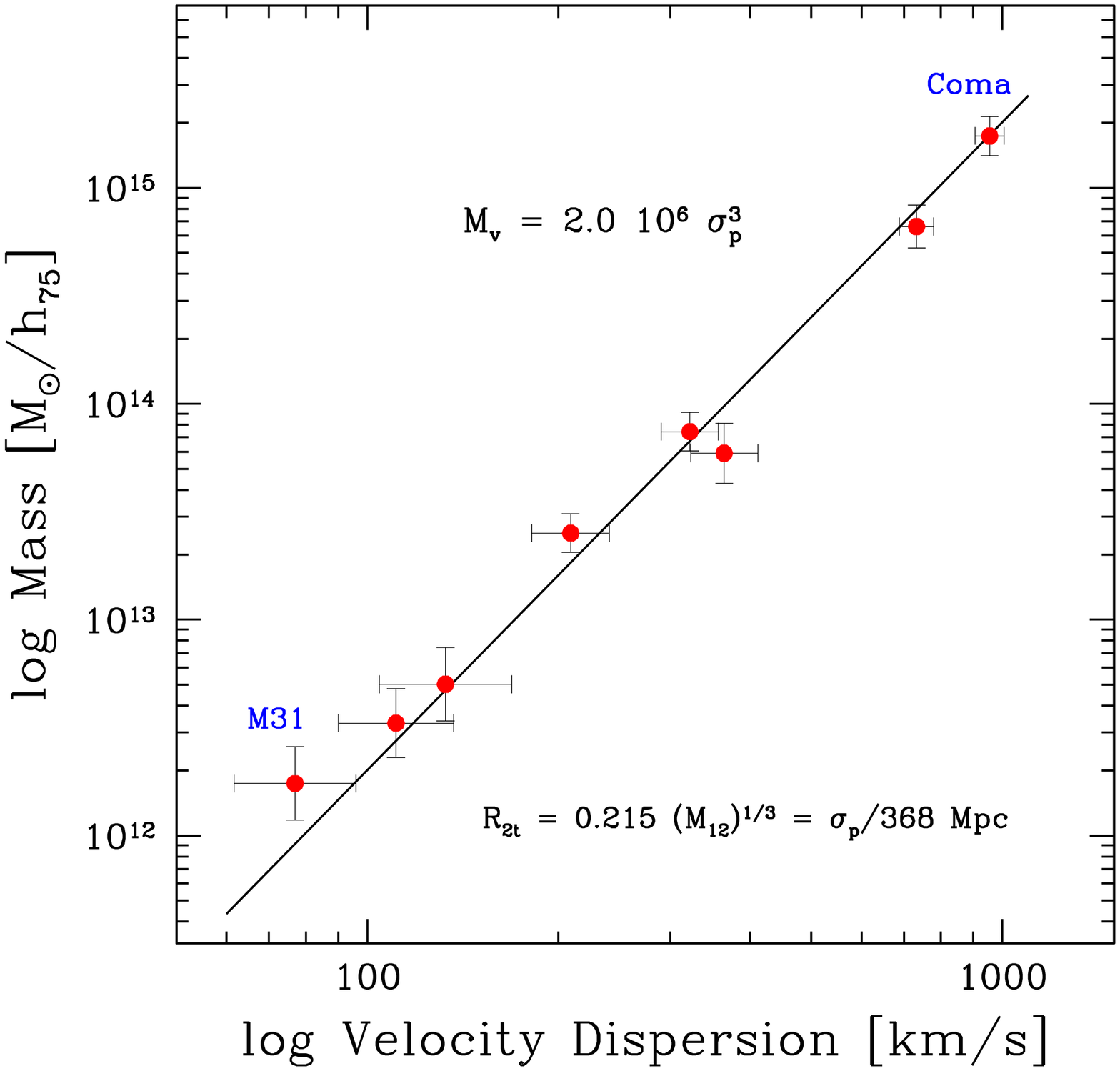} 
 \caption{Scaling relations between $R_{2t}$, $\sigma_p$ and $M_v$.   In each case, the slope is given by the theoretical expectation and the only free parameter is the scale zero point.}
\end{center}
\end{figure}

\bigskip
The second paper (Tully 2015$b$) used these scaling relations to build a group catalog.  The input sample was the 2MASS Redshift Survey complete to $K_s = 11.75$ mag (Huchra et al. 2012) with 43,038 galaxies.  An assumptions was needed regarding the relationship between the mass of a group and the luminosity of constituent galaxies.  A formulation was motivated that ran linearly with log mass from $M_v/L_K=40$ at $10^{12}~M_{\odot}$ to $M_v/L_K=120$ at $10^{15}~M_{\odot}$.

A group catalog could be built with these pieces.  Using redshifts for distance, the most intrinsically luminous galaxy in the catalog was identified, its inferred mass was calculated, whence its expectation second turnaround radius and associated velocity dispersion.  Galaxies within this radius and twice the velocity dispersion were linked as group members.  The added candidates increased the joint luminosity, hence the associated parameters so the search was renewed for additional members.  In this way, the entire catalog was processed.

The range of optimal validity of the group catalog is 3,000$-$10,000~km/s.  Beyond 10,000~km/s the selection function has large uncertainties.  Nearer than 3,000~km/s the 2MASS catalog is not optimal because of the loss of low surface brightness systems.  In the range 3,000$-$10,000~km/s there are 24,044 galaxies in the catalog.  They are assigned to 3461 groups of 2 or more and 10,145 singles.  A halo mass function was presented that should have validity over the mass range $6\times10^{12}$ to $5\times10^{14}~M_{\odot}$.

\bigskip\noindent
Support for this program has been provided by the US National Science Foundation, the NASA Astrophysics Data Analysis Program, and awards from the Space Telescope Science Institute in connection with observations with Hubble Space Telescope.  Additional observations were made with the Canada-France Hawaii and Subaru telescopes.

\end{document}